\documentclass[aps,prl,preprint,groupedaddress,nofootinbib,showkeys]{revtex4}
\usepackage{amsmath,amssymb,graphicx}
\begin{document}
\preprint{\parbox[b]{1in}{ \hbox{\tt PNUTP-16/A01} }}
%\preprint{}

\title{Composite (pseudo) scalar contributions to muon $g-2$}


\author{Deog Ki Hong}
\email[]{dkhong@pusan.ac.kr}
\author{Du Hwan Kim}
\affiliation{Department of
Physics,   Pusan National University,
             Busan 46241, Korea}


\date{\today}

\begin{abstract}
	We have calculated the composite (pseudo) scalar contributions to the anomalous magnetic moment of muons in models of walking technicolor. By the axial or scale anomaly the light scalars such as techni-dilaton, techni-pions or techni-eta have anomalous couplings to two-photons, which make them natural candidates for the recent 750 GeV resonance excess, observed at LHC. Due to the anomalous couplings, their contributions to muon $(g-2)$ are less suppressed and might explain the current deviation in muon $(g-2)$ measurements from theory.

\end{abstract}


\maketitle

\subsection{Introduction}
After discovery of Higgs boson~\cite{Aad:2012tfa,Chatrchyan:2012xdj}, signals for new physics beyond standard model (BSM) have been intensively searched at LHC. Very recently both the ATLAS and CMS group have observed an excess at 750 GeV with the local significance by about 3 $\sigma$ in the diphoton channel at the 13 TeV LHC~\cite{atlas,CMS:2015dxe}, which, if confirmed, will be a genuine direct signal for new physics at colliders. There have been since proposed numerous models of BSM to explain this single resonance. In establishing correct models of BSM, it will be therefore desired to constrain those proposed models, if possible, from the precision measurement of low energy physics~\cite{Delaunay:2016zmu,Frugiuele:2016rii,Goertz:2015nkp}, which often severely constrain BSM models, otherwise difficult to be ruled out at colliders.
 
It is well known that the standard model (SM) estimation of the anomalous magnetic moment of the muon has quite a significant deviation from  the experiments, which might be due to a new physics beyond standard model.
 Recent measurements of the anomalous magnetic moment of the  muon~\cite{Bennett:2006fi}, performed at the Brookhaven National Laboratory (BNL), find 
\begin{equation}
a_{\mu}=11 659 208.0(5.4)(3.3)\times10^{-10}\,,
\end{equation}
which deviates by $3.2~\sigma$  above the current SM estimate, based on $e^{+}e^{-}$ hadronic cross sections~\cite{Bijnens:2007pz,Jegerlehner:2009ry}. An improved muon $(g-2)$ experiment is approved and under construction at the Fermilab to achieve a precision of $0.14$ ppm~\cite{Miller:2007kk,Gray:2015qna}, which will move the deviation, if persistent, to 5~$\sigma$.

In this paper, we estimate the new physics contributions to the anomalous magnetic moment of muons to see if it naturally fixes the current deviation, provided that the recent excess at $750~{\rm GeV}$ at ATLAS and CMS is due to the scalar or pseudo-scalar resonances, predicted in the models of new strong dynamics such as walking technicolor~\cite{Molinaro:2015cwg,Matsuzaki:2015che} or models of composite axions~\cite{Barrie:2016ntq}.

\subsection{Candidates for 750 GeV resonance }
The anomalous magnetic moment of muons is one of a few physical observables that are measured so precisely, with an accuracy of parts per million (ppm). It is therefore quite sensitive to new physics at TeV, since generically the new physics contribution to the anomalous magnetic moment of muons is given by the dimensional analysis as 
\begin{equation}
a_{\mu}^{\rm {NP}}\sim \frac{\alpha_{\rm em}}{\pi}\left(\frac{m}{M_{\rm NP}}\right)^2\sim10^{-10}\left(\frac{1\,{\rm TeV}}{M_{\rm NP}}\right)^2\,,
\end{equation}
where $m$ is the muon mass and $M_{\rm NP}$ is a typical scale of new physics. 
Generically the new physics contribution is well within the experimental accuracy and thus might explain the current $3.2~\sigma$ deviation~\cite{Jegerlehner:2009ry} if $M_{\rm NP}$ is not too higher than $1~{\rm TeV}$. Indeed the new physics contribution is well studied up to two-loops for the weakly interacting new particles to exclude certain parameter regions in some extension of the standard model such as MSSM or simplified models~\cite{Cho:2001nfa,Queiroz:2014zfa}. In this paper we focus on strong dynamics extension of the standard model, especially the walking technicolor (WTC) models which break dynamically not only the electroweak symmetry but also the approximate scale symmetry, introduced to accommodate the constraints from  
the electroweak precision measurements~\cite{Holdom:1981rm,Yamawaki:1985zg,Hong:2004td}. 

By the hypothesis of partially conserved dilatation currents (PCDC)  among the spin-0 excitations 
of WTC the lightest one should be the techni-dilaton, which is a pseudo Nambu-Goldstone boson, associated with  the spontaneous broken scale symmetry~\cite{Yamawaki:1985zg,Bando:1986bg,Hong:2013eta,Choi:2012kx}. Since PCDC assumes the techni-dilaton saturates the matrix elements of dilatation current at low energy, we have from the trace anomaly~\cite{Hong:2013eta,Choi:2012kx}
\begin{equation}
F_{\rm D}^2M_D^2 \sim m_{\rm TC}^4,
\end{equation}
where $F_D$ and $M_D$ are the dilaton decay constant and mass, respectively, and the trace anomaly is given mostly by the dynamical mass of techni-fermions~\cite{Miransky:1989qc}, $m_{\rm TC}$, which characterizes the IR scale of WTC, about 1 TeV.  Having the theory very near the quasi infrared (IR) fixed point, one can separate widely the ultra-violet (UV) scale from the IR scale of WTC or $F_{\rm D}\gg m_{\rm TC}$ to have a light dilaton, $M_{\rm D}\sim m^2_{\rm TC}/F_{\rm D}\ll m_{\rm TC}\sim 1~{\rm TeV}$~\cite{Hong:2013eta,Choi:2012kx}. 
It is therefore quite natural to interpret the 125 GeV boson, discovered at LHC~\cite{Aad:2012tfa,Chatrchyan:2012xdj} as the techni-dilaton, if WTC is responsible for electroweak symmetry breaking that describes the BSM physics. Compared to the standard model Higgs, the techni-dilaton couples to gluons more strongly but to SM fermions more weakly. Hence, with current LHC data the techni-dilaton is still a viable interpretation of the 125 GeV boson~\cite{Matsuzaki:2015sya}. On the other hand, if WTC is not so extremely conformal, the IR and UV scales are not widely separated and  the techni-dilaton mass will not be much smaller than the typical IR scale of WTC, $m_{\rm TC}\sim 1~{\rm TeV}$, but it should be still the lightest one in the spectrum by PCDC, though it lies close to other scalar excitations such as the composite Higgs. In this case the 125 GeV boson may be interpreted as the composite Higgs of WTC~\cite{Hong:2004td}, since  it can be very light due to the top-quark loop corrections~\cite{Foadi:2012bb}, and then the 750 GeV resonance may be interpreted as the techni-dilaton of WTC that decays into two photons by the trace anomaly.

Being Nambu-Goldstone bosons, pseudo scalars that are associated with spontaneously broken chiral symmetry of techni-fermions in WTC are generically also light, though they are strongly coupled. If the 750 GeV resonance is a pseudo scalar, it may be interpreted as either techni-pion~\cite{Matsuzaki:2015che} or techni-eta in WTC~\cite{Franceschini:2015kwy} or a composite axion~\cite{Barrie:2016ntq}, which decay into two photons by the Adler-Bell-Jackiw anomaly~\cite{Adler:1969gk,Bell:1969ts}.

\subsection{New scalars contributions to muon $(g-2)$}
\begin{figure}[tbh]
	\includegraphics[width=0.28\textwidth]{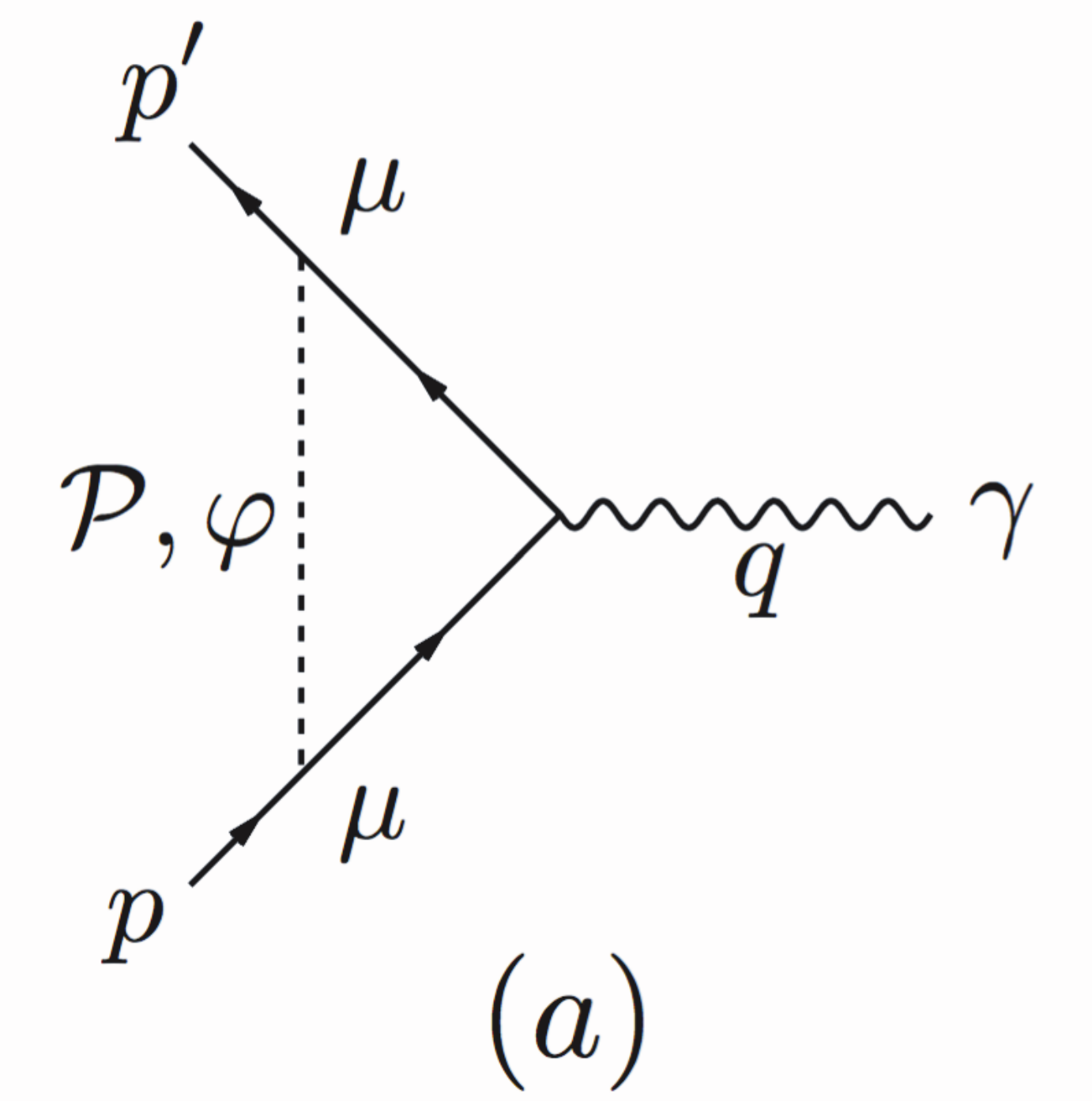}\vskip 0,1in\hskip 0.1in
	\includegraphics[width=0.6\textwidth]{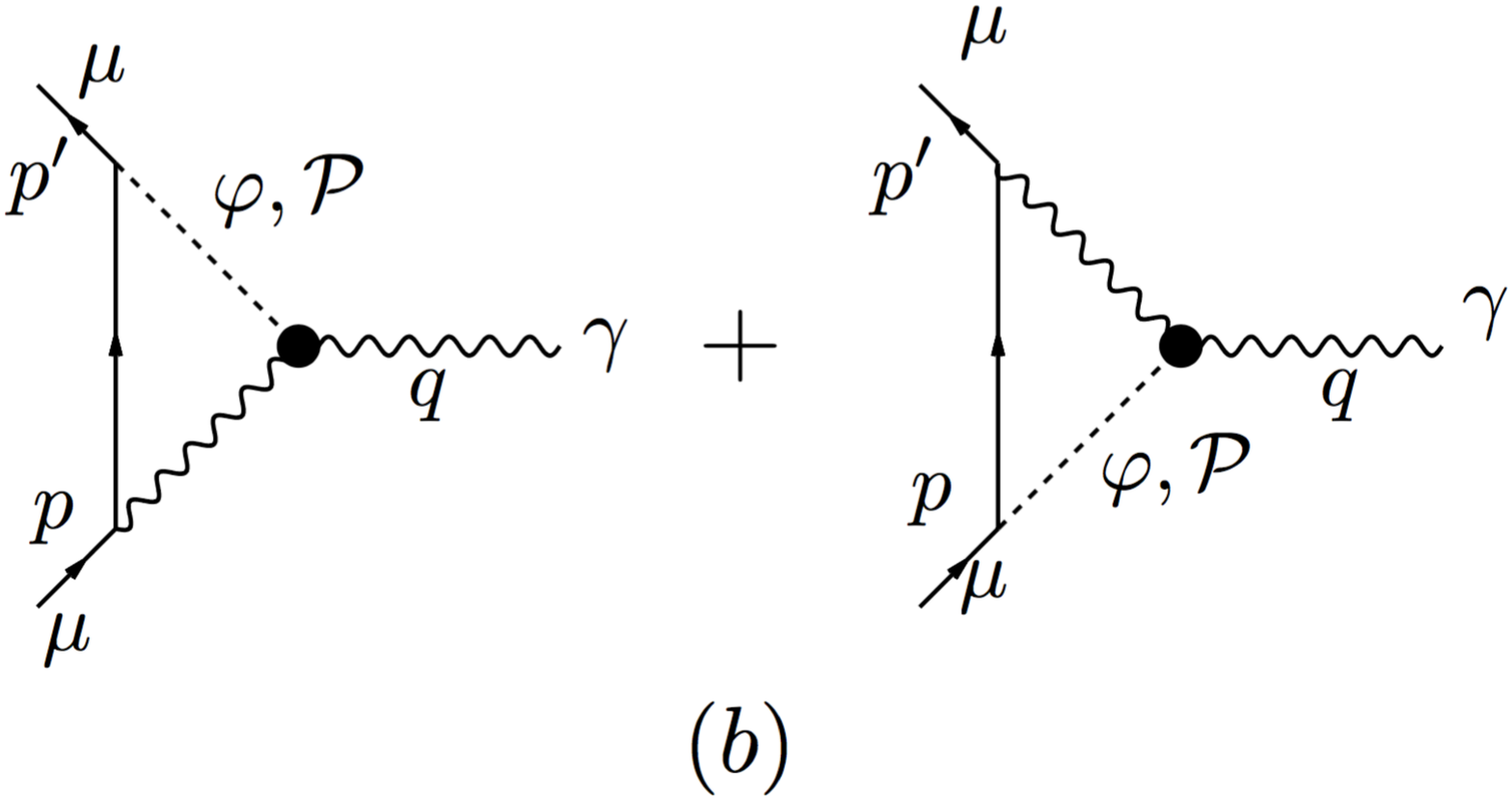}
	\caption{One-loop corrections to the anomalous magnetic moment of muons. The dotted line denotes either  the techni-dilaton field ($\varphi$) or  a pseudo scalar ${\cal P}$, the neutral techni-pion or techni-eta (or composite axion): (a) A diagram similar to the one-loop SM Higgs contribution.  (b) One-loop diagram due to anomalous couplings of (pseudo) scalar to photons, denoted as a blob.}
	\label{fig1}
	 \end{figure}

The low energy interaction Lagrangian density, relevant for our discussions on the muon $(g-2)$, is given as~\footnote{The couplings of composite (pseudo) scalars with other standard model particles besides muons and photons will be  relevant only for two or higher loop contributions to muon $g-2$. For instance the techni-dilaton coupling to two-photons through the W boson loop will contribute to muon $g-2$ at two-loop, as in the case of Higgs contributions, but further suppressed by $(v_{\rm ew}/F_D)^2$ and thus much smaller than our one-loop results.}
\begin{equation}
{\cal L}_{\rm int}=-{\bar \psi}\left(g_D\varphi+i\gamma_5g_A{\cal P}\right)\psi+e^2\frac{c_D}{4F_{\rm D}}\varphi F_{\mu\nu}F^{\mu\nu}+e^2\frac{c_A}{4F_{\rm A}}{\cal P}F_{\mu}{\tilde F}^{\mu\nu}\,,
\label{eff}
\end{equation}
where the techni-dilaton ($\varphi$) coupling to the muon field, denoted as $\psi$, $g_D=(3-\gamma_m)m/F_{\rm D}$ with the anomalous dimension of the techni-fermion bilinear, $\gamma_m\approx1$. The pseudo-scalar coupling $g_A$, induced by the extended technicolor (ETC) is given as 
$\sqrt{3}m/(2F_{\pi})$~\cite{Jia:2012kd}. 
$F_{\mu\nu}=\partial_{\mu}A_{\nu}-\partial_{\nu}A_{\mu}$, the field strength tensor of the photon field $A_{\mu}$ with the electric charge $e$ and $\tilde F_{\mu\nu}$ is its dual. The two-photon coupling of techni-dilaton, $c_D$, or pseudo-scalar (${\cal P}$), $c_A$, is in general the momentum-dependent anomalous form factor but can be regarded as a constant in the effective theory, which is determined by the UV physics anomaly~\footnote{The exact form of the anomalous form factor is difficult to calculate due to its non-perturbative nature.   
As in the QCD corrections to the light-by-light contribution to muon (g-2), one may approximate, to correctly reproduce its asymptotic UV behavior similar to the Lepage-Brodsky formula in QCD, 
the anomalous form factor by a single (techni) vector-meson  $F_{i\gamma\gamma}(q^2,Q^2)\approx\,C_iM_V^2/(Q^2+M_V^2)$~\cite{Jegerlehner:2009ry} or an infinite tower of (techni) vector mesons in holographic models~\cite{Hong:2009zw}. However, since the form factor will significantly differ from the constant approximation only for the internal momentum bigger than the vector meson mass, $Q^2\gtrsim M_V^2$ and the loop diagram Fig.~\ref{fig1} (b) is dominant by the momentum smaller than the (pseudo) scalar mass, $M_i^2$, the error that we are making is about $M_i^2/M_V^2\approx (M_i/4\pi F_i)^2\lesssim0.25$, if $M_V\gtrsim1.5~{\rm TeV}$.
}.

At one-loop the (pseudo) scalar contributions to the anomalous magnetic moment of muons  consists of two pieces (see Fig.~\ref{fig1}). The diagram in Fig.~\ref{fig1}(a), which is same as the one-loop Higgs contribution except the couplings and mass, gives 
\begin{equation}
a_{\mu}^{\rm NP(a)}\simeq\frac{g_i^2}{8\pi^2}\frac{m^2}{M_i^2}\ln\left(\frac{M_i^2}{m^2}\right)\,,
\label{higgs}
\end{equation}
where $i$ denotes either $D$ for the techni-dilaton or $A$ for the pseudo scalar fields. 
From the anomalous coupling diagram, Fig.~\ref{fig1}(b), we find with ${\bar g_i}=g_i\cdot F_i/m$
\begin{equation}
a_{\mu}^{{\rm NP}(b)}\simeq \frac{\alpha_{\rm em}}{2\pi}\,{\bar g_i}c_i \frac{m^2}{F_i^2}\cdot\ln\left(\frac{16\pi^2 F_i^2}{M_i^2}\right)\,\sim 10^{-9}\left(\frac{{\bar g_i}c_i}{2.5}\right)\cdot\left(\frac{0.5~{\rm TeV}}{F_i}\right)^2\,.
\label{anomaly}
\end{equation}
where we have taken $4\pi F_{i}$ as the UV cutoff of the effective interactions in Eq.~(\ref{eff}), following the naive dimensional analysis~\cite{Manohar:1983md}.

\subsection{Results and Discussion}
For the contribution from the diagram in Fig.~{\ref{fig1}(a), Eq.~(\ref{higgs}), which is doubly suppressed by $(m/F_i)^2$ and $(M_H/M_i)^2$, is more suppressed than the one-loop SM Higgs contribution, $a_{\mu}^{(2){\rm EW(H)}}<5\times 10^{-14}$:
\begin{equation}
a_{\mu}^{\rm NP(a)}\approx a_{\mu}^{(2){\rm EW(H)}}\cdot\left(\frac{M_H}{M_i}\right)^2\cdot\left(\frac{v_{\rm EW}}{F_i}\right)^2\,,
\end{equation}
where the vacuum expectation value of Higgs, $v_{\rm EW}=246~{\rm GeV}$. $M_H=125~{\rm GeV}$ is the Higgs mass and $M_i$ ($F_i$) is the mass (decay constant) of either the techni-dilaton or a pseudo scalar. On the other hand, since  the one-loop contribution, Eq.~(\ref{anomaly}), from the anomalous coupling, the diagram in Fig.~{\ref{fig1}(b) is singly suppressed by $(m/F_i)^2$, compared to the one-loop QED contribution, it may be comparable to the current $3.2\sigma$ deviation, $\Delta a_{\mu}=a_{\mu}^{\rm exp}-a_{\mu}^{\rm th}\approx (290\pm90)\times 10^{-11}$~\cite{Jegerlehner:2009ry}. Indeed, for ${\bar g_i}c_i=2.5$ and $F_i=0.5~{\rm TeV}$, we find that the new physics contribution Eq.~(\ref{anomaly})  is of the order of $\Delta a_{\mu}$. However, if $F_i$ is much bigger than $0.5~{\rm TeV}$ or the product of the Yukawa and diphotons couplings, ${\bar g_i}c_i$, of the (pseudo) scalar is too small, the muon anomaly may not be explained in models of WTC. 

In models of WTC, the anomalous couplings $c_i$'s are generically too small to account for the muon anomaly, since they are suppressed by the techni-fermion loop factors~\footnote{In the case of one-family WTC model, the neutral techni-pion is conjectured to the 750 GeV resonance~\cite{Matsuzaki:2015che}. In this model, the number of technicolor $N_{\rm TC}=3$, the techni-pion decay constant $F_{\pi}=123~{\rm GeV}$, and $c_A{\bar g_A}=1/(2\pi)^2$, which gives $a_{\mu}^{\rm NP}\approx 6\times 10^{-11}$.}. 
One could add to this model an extra techni-lepton of an electric charge $qe$, which is electrically charged but QCD-color neutral, to enhance the diphoton coupling.  For a minimal WTC model~\cite{Hong:2004td,Sannino:2004qp}, which has one techni-fermion doublet of the symmetric second-rank tensor with the electric charge $(q+1, q)$ will give the anomalous coupling $c_D=N_{\rm TC}(N_{\rm TC}+1)(2q^2+2q+1)/(12\pi^2)$ from the one-loop QED beta function.   For $q=3$ we get $c_D=2.53$, if $N_{\rm TC}=3$, but the beta function is still perturbative,  $\beta_{\rm QED}(e)/e<1$~\cite{Goertz:2015nkp}~\footnote{However, if we introduce too many electrically-charged techni-fermions or a techni-lepton with too large electric charge,  QED or $U(1)_Y$ might develop a Landau pole at much below the Planck scale.}. Similarly for the case of pseudo-scalars, one can enhance the anomalous coupling $c_A$ by introducing techni-fermions with large axial charges. %One might then be able to explain simultaneously the muon anomaly and the 750 GeV excess at LHC.

To conclude, we have calculated the one-loop contributions to the anomalous magnetic moment of muons in models of walking technicolor, which contain generically a light techni-dilaton,  techni-pions or techni-eta. At one-loop the diagrams that involve the anomalous coupling of (pseudo) scalar to two-photons is suppressed only by a single power of muon mass squared , $(m/F_i)^2$, compared the one-loop QED contribution, where $F_i$ is the decay constant of either techni-dilaton or techni-pion (eta), roughly of the order of the ultra-violet scale of the effective interaction, Eq. (\ref{eff}). We find for $F_i\sim 0.5~{\rm TeV}$ and the anomalous coupling to diphoton $c_i={\cal O}(1)$ the one-loop contribution of WTC is comparable to the current $3.2\sigma$ deviation and thus may explain the deviation in the anomalous magnetic moment of muon. However, the anomalous couplings are generically small in models of WTC, since they are suppressed by the loop factors, unless one introduces extra techni-fermions with large electric charges or axial charges. In this case the 125 GeV Higgs is the composite Higgs and the 750 GeV resonance is the techni-dilaton or the techni-pion or techni-eta in the scenario of WTC.

\subsection{Acknowledgements}
This research was supported by Basic Science Research Program through the National Research Foundation of Korea (NRF) funded by the Ministry of Education (NRF-2013R1A1A2011933). 
\subsubsection{}

\vskip 0.5in
{\bf  Note added:} After this paper is finished, it appeared the paper~\cite{Baek:2016uqf} which studies a similar problem but for the weakly interacting new (pseudo) scalars.

\begin{thebibliography}{99}
%\cite{Aad:2012tfa,Chatrchyan:2012xdj}
\bibitem{Aad:2012tfa}
  G.~Aad {\it et al.} [ATLAS Collaboration],
  %``Observation of a new particle in the search for the Standard Model Higgs boson with the ATLAS detector at the LHC,''
  Phys.\ Lett.\ B {\bf 716} (2012) 1
  doi:10.1016/j.physletb.2012.08.020
  [arXiv:1207.7214 [hep-ex]].
  %%CITATION = doi:10.1016/j.physletb.2012.08.020;%%
  %5527 citations counted in INSPIRE as of 19 Feb 2016
  %\cite{Chatrchyan:2012xdj}
\bibitem{Chatrchyan:2012xdj}
  S.~Chatrchyan {\it et al.} [CMS Collaboration],
  %``Observation of a new boson at a mass of 125 GeV with the CMS experiment at the LHC,''
  Phys.\ Lett.\ B {\bf 716} (2012) 30
  doi:10.1016/j.physletb.2012.08.021
  [arXiv:1207.7235 [hep-ex]].
  %%CITATION = doi:10.1016/j.physletb.2012.08.021;%%
  %5417 citations counted in INSPIRE as of 19 Feb 2016
%\cite{atlas,CMS:2015dxe}
\bibitem{atlas}
  The ATLAS collaboration,
  %``Search for resonances decaying to photon pairs in 3.2 fb$^{-1}$ of $pp$ collisions at $\sqrt{s}$ = 13 TeV with the ATLAS detector,''
  ATLAS-CONF-2015-081.
  %%CITATION = ATLAS-CONF-2015-081;%%
  %199 citations counted in INSPIRE as of 19 Feb 2016
%\cite{CMS:2015dxe}
\bibitem{CMS:2015dxe}
  CMS Collaboration [CMS Collaboration],
  %``Search for new physics in high mass diphoton events in proton-proton
  collisions at 13TeV,''
  CMS-PAS-EXO-15-004.
  %%CITATION = CMS-PAS-EXO-15-004;%%
  %192 citations counted in INSPIRE as of 19 Feb 2016
  
  
%\cite{Delaunay:2016zmu,Frugiuele:2016rii}
\bibitem{Delaunay:2016zmu}
  C.~Delaunay and Y.~Soreq,
  %``Probing New Physics with Isotope Shift Spectroscopy,''
  arXiv:1602.04838 [hep-ph].
  %%CITATION = ARXIV:1602.04838;%%

%\cite{Frugiuele:2016rii}
\bibitem{Frugiuele:2016rii}
  C.~Frugiuele, E.~Fuchs, G.~Perez and M.~Schlaffer,
  %``Atomic probes of new physics,''
  arXiv:1602.04822 [hep-ph].
  %%CITATION = ARXIV:1602.04822;%%

%\cite{Goertz:2015nkp}
\bibitem{Goertz:2015nkp}
  F.~Goertz, J.~F.~Kamenik, A.~Katz and M.~Nardecchia,
  %``Indirect Constraints on the Scalar Di-Photon Resonance at the LHC,''
  arXiv:1512.08500 [hep-ph].
  %%CITATION = ARXIV:1512.08500;%%
  %65 citations counted in INSPIRE as of 08 Apr 2016

%\cite{Bennett:2006fi}
\bibitem{Bennett:2006fi}
  G.~W.~Bennett {\it et al.}  [Muon G-2 Collaboration],
  %``Final report of the muon E821 anomalous magnetic moment measurement at
  %BNL,''
  Phys.\ Rev.\  D {\bf 73}, 072003 (2006).
  %[arXiv:hep-ex/0602035].
  %%CITATION = PHRVA,D73,072003;%%

%\cite{Bijnens:2007pz}
\bibitem{Bijnens:2007pz}
 For  recent reviews, see  J.~Bijnens and J.~Prades,
  %``The hadronic light-by-light contribution to the muon anomalous magnetic
  %moment: Where do we stand?,''
  Mod.\ Phys.\ Lett.\  A {\bf 22}, 767 (2007)
  [arXiv:hep-ph/0702170];
  %%CITATION = MPLAE,A22,767;%%  
%\cite{Prades:2009tw}
%\bibitem{Prades:2009tw}
  J.~Prades, E.~de Rafael and A.~Vainshtein,
  %``The Hadronic Light-by-Light Scattering Contribution to the Muon and Electron Anomalous Magnetic Moments,''
  Adv.\ Ser.\ Direct.\ High Energy Phys.\  {\bf 20} (2009) 303
 % doi:10.1142/9789814271844_0009
  [arXiv:0901.0306 [hep-ph]];
  and the reference below~\cite{Jegerlehner:2009ry}.
  %%CITATION = doi:10.1142/9789814271844_0009;%%
  %178 citations counted in INSPIRE as of 19 Feb 2016%\cite{Jegerlehner:2009ry}
%\cite{Jegerlehner:2009ry}
\bibitem{Jegerlehner:2009ry}
  F.~Jegerlehner and A.~Nyffeler,
  %``The Muon g-2,''
  Phys.\ Rept.\  {\bf 477} (2009) 1
 % doi:10.1016/j.physrep.2009.04.003
  [arXiv:0902.3360 [hep-ph]].
  %%CITATION = doi:10.1016/j.physrep.2009.04.003;%%
  %516 citations counted in INSPIRE as of 19 Feb 2016

%\cite{Miller:2007kk}
\bibitem{Miller:2007kk}
  J.~P.~Miller, E.~de Rafael and B.~L.~Roberts,
  %``Muon g-2: Review of Theory and Experiment,''
  Rept.\ Prog.\ Phys.\  {\bf 70}, 795 (2007)
  [arXiv:hep-ph/0703049].
  %%CITATION = RPPHA,70,795;%%
%\cite{Gray:2015qna}
\bibitem{Gray:2015qna}
  F.~Gray [Muon g-2 Collaboration],
  %``Muon g-2 Experiment at Fermilab,''
  arXiv:1510.00346 [physics.ins-det].
  %%CITATION = ARXIV:1510.00346;%%
  %2 citations counted in INSPIRE as of 19 Feb 2016

%\cite{Molinaro:2015cwg,Matsuzaki:2015che}
\bibitem{Molinaro:2015cwg}
  E.~Molinaro, F.~Sannino and N.~Vignaroli,
  %``Minimal Composite Dynamics versus Axion Origin of the Diphoton excess,''
  arXiv:1512.05334 [hep-ph].
  %%CITATION = ARXIV:1512.05334;%%
  %135 citations counted in INSPIRE as of 19 Feb 2016  
  %\cite{Matsuzaki:2015che}
\bibitem{Matsuzaki:2015che}
  S.~Matsuzaki and K.~Yamawaki,
  %``750 GeV Diphoton Signal from One-Family Walking Technipion,''
  arXiv:1512.05564 [hep-ph].
  %%CITATION = ARXIV:1512.05564;%%
  %105 citations counted in INSPIRE as of 19 Feb 2016
 %\cite{Barrie:2016ntq}
\bibitem{Barrie:2016ntq}
  N.~D.~Barrie, A.~Kobakhidze, M.~Talia and L.~Wu,
  %``750 GeV Composite Axion as the LHC Diphoton Resonance,''
  doi:10.1016/j.physletb.2016.02.010
  arXiv:1602.00475 [hep-ph].
  %%CITATION = doi:10.1016/j.physletb.2016.02.010;%%
  %3 citations counted in INSPIRE as of 19 Feb 2016 
  
%\cite{Cho:2001nfa,Queiroz:2014zfa}
\bibitem{Cho:2001nfa}
  G.~C.~Cho and K.~Hagiwara,
  %``Supersymmetric contributions to muon g-2 and the electroweak precision measurements,''
  Phys.\ Lett.\ B {\bf 514} (2001) 123
  doi:10.1016/S0370-2693(01)00815-2
  [hep-ph/0105037].
  %%CITATION = doi:10.1016/S0370-2693(01)00815-2;%%
  %33 citations counted in INSPIRE as of 19 Feb 2016
  %\cite{Queiroz:2014zfa}
\bibitem{Queiroz:2014zfa}
  F.~S.~Queiroz and W.~Shepherd,
  %``New Physics Contributions to the Muon Anomalous Magnetic Moment: A Numerical Code,''
  Phys.\ Rev.\ D {\bf 89} (2014) 9,  095024
  doi:10.1103/PhysRevD.89.095024
  [arXiv:1403.2309 [hep-ph]].
  %%CITATION = doi:10.1103/PhysRevD.89.095024;%%
  %30 citations counted in INSPIRE as of 19 Feb 2016
    
  %\cite{Holdom:1981rm,Yamawaki:1985zg,Hong:2004td}
\bibitem{Holdom:1981rm}
  B.~Holdom,
  %``Raising The Sideways Scale,''
  Phys.\ Rev.\  D {\bf 24} (1981) 1441.
  %%CITATION = PHRVA,D24,1441;%%


\bibitem{Yamawaki:1985zg}
  K.~Yamawaki, M.~Bando and K.~Matumoto,
  %``Scale Invariant TC Model And A Technidilaton,''
  Phys.\ Rev.\ Lett.\  {\bf 56}, 1335 (1986). 
  %%CITATION = PRLTA,56,1335;%%
  
  %\cite{Hong:2004td}
\bibitem{Hong:2004td}
  D.~K.~Hong, S.~D.~H.~Hsu and F.~Sannino,
  %``Composite Higgs from higher representations,''
  Phys.\ Lett.\  B {\bf 597}, 89 (2004);
  %[arXiv:hep-ph/0406200];
  %%CITATION = PHLTA,B597,89;%%
  
  %\cite{Bando:1986bg,Hong:2013eta,Choi:2012kx}
\bibitem{Bando:1986bg}
  M.~Bando, K.~i.~Matumoto and K.~Yamawaki,
  %``Technidilaton,''
  Phys.\ Lett.\ B {\bf 178} (1986) 308.
  doi:10.1016/0370-2693(86)91516-9
  %%CITATION = doi:10.1016/0370-2693(86)91516-9;%%
  %158 citations counted in INSPIRE as of 20 Feb 2016
  %\cite{Hong:2013eta}
\bibitem{Hong:2013eta}
  D.~K.~Hong,
  %``Composite Higgs and Techni-Dilaton at LHC,''
  doi:10.1142/9789814566254$_-$0020,
  arXiv:1304.7832 [hep-ph].
  %%CITATION = doi:10.1142/9789814566254_0020;%%
  %2 citations counted in INSPIRE as of 20 Feb 2016
  %\cite{Choi:2012kx}
\bibitem{Choi:2012kx}
  K.~Y.~Choi, D.~K.~Hong and S.~Matsuzaki,
  %``Analysis of techni-dilaton as a dark matter candidate,''
  JHEP {\bf 1212} (2012) 059
  doi:10.1007/JHEP12(2012)059
  [arXiv:1201.4988 [hep-ph]].
  %%CITATION = doi:10.1007/JHEP12(2012)059;%%
  %2 citations counted in INSPIRE as of 20 Feb 2016
  
  %\cite{Miransky:1989qc}
\bibitem{Miransky:1989qc}
  V.~A.~Miransky and V.~P.~Gusynin,
  %``Chiral Symmetry Breaking and Nonperturbative Scale Anomaly in Gauge Field Theories,''
  Prog.\ Theor.\ Phys.\  {\bf 81} (1989) 426.
  doi:10.1143/PTP.81.426
  %%CITATION = doi:10.1143/PTP.81.426;%%
  %52 citations counted in INSPIRE as of 20 Feb 2016
 %\cite{Matsuzaki:2015sya}
\bibitem{Matsuzaki:2015sya}
  S.~Matsuzaki and K.~Yamawaki,
  %``Walking on the ladder: 125 GeV technidilaton, or Conformal Higgs,''
  JHEP {\bf 1512} (2015) 053
  doi:10.1007/JHEP12(2015)053
  [arXiv:1508.07688 [hep-ph]].
  %%CITATION = doi:10.1007/JHEP12(2015)053;%%
  %4 citations counted in INSPIRE as of 20 Feb 2016 
 %\cite{Foadi:2012bb}
\bibitem{Foadi:2012bb}
  R.~Foadi, M.~T.~Frandsen and F.~Sannino,
  %``125 GeV Higgs boson from a not so light technicolor scalar,''
  Phys.\ Rev.\ D {\bf 87} (2013) 9,  095001
  doi:10.1103/PhysRevD.87.095001
  [arXiv:1211.1083 [hep-ph]].
  %%CITATION = doi:10.1103/PhysRevD.87.095001;%%
  %67 citations counted in INSPIRE as of 20 Feb 2016 
 %\cite{Franceschini:2015kwy}
\bibitem{Franceschini:2015kwy}
  R.~Franceschini {\it et al.},
  %``What is the gamma gamma resonance at 750 GeV?,''
  arXiv:1512.04933 [hep-ph].
  %%CITATION = ARXIV:1512.04933;%%
  %184 citations counted in INSPIRE as of 20 Feb 2016 
 
  %\cite{Adler:1969gk,Bell:1969ts}
\bibitem{Adler:1969gk}
  S.~L.~Adler,
  %``Axial vector vertex in spinor electrodynamics,''
  Phys.\ Rev.\  {\bf 177} (1969) 2426.
  doi:10.1103/PhysRev.177.2426
  %%CITATION = doi:10.1103/PhysRev.177.2426;%%
  %3148 citations counted in INSPIRE as of 20 Feb 2016
 %\cite{Bell:1969ts}
\bibitem{Bell:1969ts}
  J.~S.~Bell and R.~Jackiw,
  %``A PCAC puzzle: pi0 --> gamma gamma in the sigma model,''
  Nuovo Cim.\ A {\bf 60} (1969) 47.
  doi:10.1007/BF02823296
  %%CITATION = doi:10.1007/BF02823296;%%
  %2594 citations counted in INSPIRE as of 20 Feb 2016
  %\cite{Jia:2012kd}
\bibitem{Jia:2012kd}
  J.~Jia, S.~Matsuzaki and K.~Yamawaki,
  %``Walking technipions at the LHC,''
  Phys.\ Rev.\ D {\bf 87} (2013) 1,  016006
  doi:10.1103/PhysRevD.87.016006
  [arXiv:1207.0735 [hep-ph]].
  %%CITATION = doi:10.1103/PhysRevD.87.016006;%%
  %15 citations counted in INSPIRE as of 21 Feb 2016
  
 %\cite{Hong:2009zw}
\bibitem{Hong:2009zw}
  D.~K.~Hong and D.~Kim,
  %``Pseudo scalar contributions to light-by-light correction of muon g-2 in AdS/QCD,''
  Phys.\ Lett.\ B {\bf 680} (2009) 480
  doi:10.1016/j.physletb.2009.09.026
  [arXiv:0904.4042 [hep-ph]].
  %%CITATION = doi:10.1016/j.physletb.2009.09.026;%%
  %27 citations counted in INSPIRE as of 02 avril 2016  
 
  
 %\cite{Manohar:1983md}
\bibitem{Manohar:1983md}
  A.~Manohar and H.~Georgi,
  %``Chiral Quarks and the Nonrelativistic Quark Model,''
  Nucl.\ Phys.\ B {\bf 234} (1984) 189.
  doi:10.1016/0550-3213(84)90231-1
  %%CITATION = doi:10.1016/0550-3213(84)90231-1;%%
  %1736 citations counted in INSPIRE as of 21 Feb 2016   
  
  
  
  %\cite{Sannino:2004qp}
\bibitem{Sannino:2004qp}
  F.~Sannino and K.~Tuominen,
  %``Orientifold theory dynamics and symmetry breaking,''
  Phys.\ Rev.\ D {\bf 71} (2005) 051901
  doi:10.1103/PhysRevD.71.051901
  [hep-ph/0405209].
  %%CITATION = doi:10.1103/PhysRevD.71.051901;%%
  %343 citations counted in INSPIRE as of 08 Apr 2016
  
  
  
  %\cite{Baek:2016uqf}
\bibitem{Baek:2016uqf}
  S.~Baek and J.~h.~Park,
  %``LHC 750 GeV diphoton excess and muon $(g-2)$,''
  arXiv:1602.05588 [hep-ph].
  %%CITATION = ARXIV:1602.05588;%%
  
  
\end{thebibliography}
\end{document}